\documentclass{ws-ijmpe}
\usepackage{url}
\usepackage[super]{cite}
\usepackage{xcolor}
\usepackage[verbose,hypertexnames=false]{hyperref}
\hypersetup{colorlinks=false,allbordercolors=blue,pdfborderstyle={/S/U/W 1}}

\begin{document}

\markboth{U. M. Yanikov, V. A. Kulikov, A. M. Shirokov}{
Charged particle scattering within Efros method}

\catchline{}{}{}{}{}

\title{
Scattering of charged particles within Efros method utilizing\\ oscillator series expansion of wave functions}

\author{Ustin M. Yanikov}

\address{Department of Physics, Moscow State University, 1-2 Leninskiye Gory Street\\
Moscow, 119991,
Russia\\
yanikov-u@yandex.ru}

\author{Vasily A. Kulikov}

\address{Skobeltsyn Institute of Nuclear Physics, Moscow State University, 1-2 Leninskiye Gory Street\\
Moscow, 119234, Russia\\
kulikov@nucl-th.sinp.msu.ru}

\author{Andrey M. Shirokov}

\address{Skobeltsyn Institute of Nuclear Physics, Moscow State University, 1-2 Leninskiye Gory Street\\
Moscow, 119234, Russia\\
shirokov@nucl-th.sinp.msu.ru}

\maketitle

\begin{history}
\received{Day Month Year}
\revised{Day Month Year}
\accepted{Day Month Year}
\published{Day Month Year}
\end{history}

\begin{abstract}
We apply the version of  the Efros method utilizing oscillator expansion of wave functions to the Coulomb scattering problem 
using our recent developments of the HORSE formalism. The approach yields accurate phase shifts and cross sections with significantly reduced computational cost compared to the full HORSE method, while maintaining agreement with exact solutions. These results demonstrate the efficiency of the Efros method and its 
prospect for applications in \textit{ab initio} nuclear 
reaction calculations.
\end{abstract}



\section{Introduction}
\label{intro}

Static properties 
of light atomic nuclei are currently calculated within various \textit{ab initio} approaches which do not use any model assumptions about nuclear structure. It is especially worth noting No-Core Shell Model (NCSM)\cite{BARRETT2013131} as one of the most advanced and promising \textit{ab initio} methods which utilizes the harmonic oscillator basis.

The nuclear continuum 
spectrum has been successfully studied so far within different approaches, e.\,g., $R$-matrix method\cite{Lane} and $J$-matrix method\cite{Yamani,Yamani2001}. The formalism of Harmonic Oscillator Representation of Scattering Equations (HORSE)\cite{osti_6451278,Filip,osti_6232444,Bang1999PmatrixAJ} is a particular version of $J$-matrix method utilizing harmonic oscillator basis which 
is preferable 
for 
many-body nuclear applications. The HORSE is the discrete analogue of $P$-matrix which is the inverse of $R$-matrix.\cite{Bang1999PmatrixAJ}

Various versions of the HORSE method 
have been successfully applied to the studies of nuclear resonant 
states,\cite{PhysRevC.94.064320, PhysRevC.98.044624} photodisintegration,~\cite{PhysRevC.106.054608} three-body continuum within phenomenological cluster models\cite{Mikh,Lurie} and various  problems within the Resonating Group Model (RGM)\cite{Arickx,Kato,Lashko}. However, 
accounting for the
Coulomb interaction between charged particles within HORSE encounters difficulties due to the long-range nature of the Coulomb potential. Several methods have been suggested to 
account for Coulomb asymptotics within the oscillator expansion of wave 
functions\cite{Yamani2001,Bang1999PmatrixAJ,OKHRIMENKO1984121}, 
however, all
of them have significant 
difficulties in combining with many-body {\it ab initio} approaches like the NCSM for
studying the nuclear continuum.
Recently we have suggested\cite{YANIKOV2025100075} the method for studying
scattering of charged particles based 
on 
analysis of Coulomb matrix elements in oscillator basis performed in Ref.~\refcite{OKHRIMENKO1984121}, 
which is simple and convenient to use within the 
HORSE
formalism and is prospective for many-body applications in combination with NCSM.

Unfortunately, 
one cannot use the 
HORSE in combination with NCSM for the studies of nuclear reactions 
due to a prohibitive computational cost
as 
it 
requires calculating extremely large number of eigenstates, even for light nuclei reactions. The method proposed recently by V.~D.~Efros~\cite{Efros} is based on variational 
Hulth{\'e}n--Kohn
method~\cite{PhysRev.92.817} and can be used for 
{\it ab initio} calculations of
nuclear reactions in combination with NCSM. Recently a modified version of the Efros method
has been suggested\cite{Sharaf2025} which utilizes the oscillator expansion of
non-Coulomb scattering
wave functions developed within the HOSRE formalism and is well-adapted for the use
in combination with NCSM in many-body applications.
If one uses the full basis of 
eigenstates 
of the truncated Hamiltonian matrix in the oscillator basis 
as short-range functions (SRFs) within the 
modified
Efros method,
it appears to be
equivalent to
the 
HORSE, however, it allows accounting for 
a small
number of 
SRFs
to
obtain adequate results.
Using
a small
number of 
SRFs
not only 
simplifies calculations significantly 
but just
opens 
possibilities for studying nuclear reactions within the framework of \textit{ab initio} approaches.

In this 
contribution,
we 
combine our method for accounting for the Coulomb distortion of wave function asymptotics 
proposed in Ref.~\refcite{YANIKOV2025100075} with the 
modified 
Efros method\cite{Efros} for the studies
of charged particles scattering.
Using model problems, we
 demonstrate that this 
 approach
 provides quite accurate results even for small number of SRFs and basis truncations accessible by modern NCSM codes.

\section{Coulomb scattering within HORSE formalism}
\label{Coul_scat}

In this section, we sketch the key expressions obtained within HORSE formalism for Coulomb scattering following the approach described in Ref.~\refcite{YANIKOV2025100075}. Consider a single-channel scattering of two particles with charges $eZ_1$ and $eZ_2$. We use the conventional partial wave expansion of the wave function and describe the system in terms of partial amplitudes $u_l(k,r)$ each of which obeys the radial Schr{\"o}dinger equation with the Hamiltonian $H^l$,
\begin{equation}
    H^lu_l(k,r) = Eu_l(k,r).
    \label{Schr_eq}
\end{equation}
Here, $l$ is the orbital quantum number, $r$ is the distance between the particles, $k = \sqrt{2\mu E}/\hbar$ is the momentum, $\mu$ is the reduced mass, and $E$ is the relative motion energy. The interaction potential between the particles is represented as $V^l=V^{Nucl,\,l} + V^{Coul}$, where $V^{Nucl,\,l}$ and $V^{Coul}$ are nuclear and Coulomb potentials, respectively. The asymptotic behavior of partial amplitudes $u_l(k,r)$ is described by the superposition of regular $F_l(\eta,kr)$ and irregular $G_l(\eta,kr)$ Coulomb wave functions\cite{abramowitz1964handbook},
\begin{equation}
    u_l(k, r) \xrightarrow[r\to\infty]{} F_l(\eta, kr) + \tan{\delta_l(k)}\,G_l(\eta, kr),
    \label{u_l}
\end{equation}
where 
$\eta = \mu Z_1 Z_2 e^2 / \hbar^2 k$ is the Sommerfeld parameter and
$\delta_l(k)$ is the phase shift.

Within the HORSE formalism, we expand the partial amplitudes $u_l(k,r)$ into a series of harmonic oscillator functions $\varphi_{nl}(r)$,
\begin{equation}
    u_l(k, r) = \sum_{n=0}^{\infty}a_{nl}(k)\,\varphi_{nl}(r),
    \label{OsBExp}
\end{equation}
where
\begin{equation}
    \varphi_{nl}(r) = (-1)^n\sqrt{\frac{2 n!}{b\Gamma (n+l+\frac{3}{2})}}\left(\frac{r}{b}\right)^{\!l+1}e^{-\frac{r^2}{2b^2}}\,L_n^{l+\frac{1}{2}}\!\!\left(\frac{r^2}{b^2}\right)\!.
\end{equation}
Here, $\Gamma(x)$ is the gamma function\cite{abramowitz1964handbook}, 
$L_n^\alpha(x)$ is the associated Laguerre 
polynomial\cite{abramowitz1964handbook}, $b = \sqrt{\hbar/\mu\omega}$ is the oscillator radius, and $\omega$ is the oscillator frequency. The expansion coefficients $a_{nl}(k)$ satisfy an infinite system of linear equations,
\begin{equation}
    \sum_{n'=0}^{\infty}(H_{nn'}^l - \delta_{nn'}E)\,a_{n'l}(k) = 0,
    \label{matr_eq}
\end{equation}
where $H_{nn'}^l = T_{nn'}^l + V_{nn'}^l$ are the Hamiltonian matrix elements in the harmonic oscillator basis with $T_{nn'}^l$ and $V_{nn'}^l$ being the matrix elements of the kinetic $T^l$ and potential $V^l$ energies, respectively.

In the harmonic oscillator basis, the kinetic energy matrix $T_{nn'}^l$ is tridiagonal,
\begin{equation}
    \begin{aligned}
        T_{nn'}^l&=0, \quad |n-n'|>1, \\
        T_{nn}^l&=\frac{\hbar\omega}{2}\left(2n+l+\frac{3}{2}\right),\\
        T_{n+1,\, n}^l=T_{n,\, n+1}^l&=-\frac{\hbar\omega}{2}\sqrt{(n+1)\left(n+l+\frac{3}{2}\right)},\\
    \end{aligned}
    \label{Tnn}
\end{equation}
with its non-zero elements $T_{nn}^l$ and $T_{n,\,n\pm 1}^l$ increasing linearly with $n$ for large values of $n$. At the same time, the matrix elements of nuclear potential $V_{nn'}^{Nucl,\,l} \to 0$ as $n$ and/or $n' \to \infty$. This allows us to replace $V^{Nucl,\,l}$ 
by
the potential $\widetilde{V}^{Nucl,\,l}$ defined by its matrix in harmonic oscillator basis truncated at some N,
\begin{equation}
    \widetilde{V}_{nn'}^{Nucl,\, l}=\left\{\begin{array}{ll}
        V_{nn'}^{Nucl,\, l} & \text{for $n$ and $n' \le N$}, \\
        0 & \text{for $n$ or $n'>N$}.
    \end{array}\right .
    \label{trunc_pot}
\end{equation}

The matrix elements of Coulomb potential $V^{Coul,\,l}_{nn'}$, however, decrease slowly along the main diagonal\cite{OKHRIMENKO1984121} and should be accounted for 
at large values of radial quantum numbers $n$.
In the Kiev group study\cite{OKHRIMENKO1984121}, it was stated that for large values of $n$, the asymptotic 
coefficients $a_{nl}^{as}(k) \equiv a_{nl}(k)$ ($n \ge N$) 
in the expansion~\eqref{OsBExp}
fit the three-term recurrence relation (TRR)
\begin{equation}
    T_{n,\, n-1}^l a_{n-1,\,l}^{as}(k)+(T_{nn}^l+V_{nn}^{ad,\,l}-E)a_{nl}^{as}(k)+T_{n,\, n+1}^l a_{n+1,\, l}^{as}(k)=0,
    \label{TRR}
\end{equation}
where the additional Coulomb term
\begin{equation}
    V_{nn}^{ad,\,l} = \hbar\omega\frac{\eta kb}{\sqrt{4n+2l+3}}.
\end{equation}
The calculations, however, demonstrate that TRR~\eqref{TRR} remains highly accurate even for small $n \sim 0$.\cite{YANIKOV2025100075} As it was shown in Ref.~\refcite{YANIKOV2025100075}, this fact allows one to replace $V^{Coul}$ and $T^l$ 
by the
operators $\widetilde{V}^{Coul,\,l}$ and $\widetilde{T}^l$, respectively, 
defined by their
matrices in oscillator basis,
\begin{align}
    \widetilde{V}_{nn'}^{Coul,\, l}&=\left\{\begin{array}{ll}
        V_{nn'}^{Coul} & \text{for $n$ and $n' \le N$}, \\
        0 & \text{for $n$ or $n'>N$},
    \end{array}\right .
    \label{trunc_pot_coul}
    \\
    \widetilde{T}_{nn'}^l&=\left\{\begin{array}{ll}
        T_{nn'}^l & \text{for $n$ and $n' \le N$}, \\
        T_{nn'}^l + V_{nn}^{ad,\,l} \delta_{nn'} & \text{for $n$ or $n'>N$}.
    \end{array}\right .
    \label{mod_kinetic}
\end{align}
The conventional HORSE formulas are then applied to the problem with Hamiltonian $\widetilde{H}^l = \widetilde{V}^{Nucl,\,l} + \widetilde{V}^{Coul,\,l} + \widetilde{T}^l$ to obtain the phase shift $\delta_l(k)$ (see Ref.~\refcite{YANIKOV2025100075} for details).

The asymptotic coefficients $a_{nl}^{as}(k)$ can be expressed as a superposition of two linearly independent solutions of TRR~\eqref{TRR}, $S_{nl}(k)$ and $C_{nl}(k)$,
\begin{equation}
    a_{nl}^{as}(k) = S_{nl}(k) + \tan{\delta_l(k)} \,C_{nl}(k).
\end{equation}
To satisfy the asymptotics~\eqref{u_l} of functions $u_l(k,r)$, we should define the solutions $S_{nl}(k)$ and $C_{nl}(k)$ so that
\begin{subequations}
    \label{Sn_Cn}
    \begin{alignat}{2}
        \sum_{n=0}^{\infty}S_{nl}(k)\,\varphi_{nl}(r)&=F_l(\eta, kr),
        \label{Sn}
        \\
        \sum_{n=0}^{\infty}C_{nl}(k)\,\varphi_{nl}(r)&=\widetilde{G}_l(\eta, kr) \xrightarrow[r\to\infty]{}G_l(\eta,kr),
        \label{Cn}
    \end{alignat}
\end{subequations}
where $\widetilde{G}_l(\eta,kr)$ is some regular function which fits an inhomogeneous Schr{\"o}dinger equation\cite{Bang1999PmatrixAJ,Zaitsev1998TrueMS}.

At large $n$, harmonic oscillator functions $\varphi_{nl}(r)$ have a delta-like behaviour in the vicinity of classical turning point $r_{turn}=\nu_{nl} b$, where $\nu_{nl} = \sqrt{4n+2l+3}$\cite{Zaitsev1998TrueMS}. Using this property, we obtain the asymptotic expressions for the coefficients $S_{nl}(k)$ and $C_{nl}(k)$\cite{YANIKOV2025100075},
\begin{subequations}
    \begin{alignat}{2}
        S_{nl}(k) \xrightarrow[n \to \infty]{} \sqrt{\frac{2b}{\nu_{nl}}}F_l(\eta, \nu kb),
        \label{sn_as}
        \\
        C_{nl}(k) \xrightarrow[n \to \infty]{} \sqrt{\frac{2b}{\nu_{nl}}}G_l(\eta, \nu kb).
        \label{cn_as}
    \end{alignat}
\end{subequations}

The high accuracy of TRR~\eqref{TRR} allows us to reproduce the coefficients $S_{nl}(k)$ and $C_{nl}(k)$\cite{YANIKOV2025100075}. For some starting value $n_s$, the coefficients $S_{n_s+2,\,l}(k)$ and $S_{n_s+1,\,l}(k)$ are obtained from the asymptotic formulas~\eqref{Sn}. The coefficients $S_{nl}(k)$ ($n=0,\dots,n_s$) are then calculated according to TRR~\eqref{TRR}. The same procedure is used to obtain the coefficients $C_{nl}(k)$ ($n=0,\dots,n_s$).

\section{Efros method}
\label{Efros_method}

The approach to finding solutions $S_{nl}(k)$ and $C_{nl}(k)$ of TRR~\eqref{TRR} described in Sec.~\ref{Coul_scat} allows us to apply Efros method\cite{Efros,Sharaf2025} to Coulomb scattering problem. Suppose that solution of the Schr{\"o}dinger equation~\eqref{Schr_eq} has the following form,
\begin{equation}
    \widetilde{u}_l(k,r) = F_l(\eta,kr) + \tan\delta_l(k)\,\widetilde{G}_l(\eta,kr) + \sum_{n=0}^{v-1}b_{nl}(k)\beta_{nl}(r),
    \label{Efros_wave_func}
\end{equation}
where $\beta_{nl}(r)$ ($n=0,\dots,v-1$) is the set of $v$ linearly independent 
SRFs.
The equation~\eqref{Efros_wave_func} 
includes
$(v+1)$ unknowns: $v$ coefficients $b_{nl}(k)$ and $\tan\delta_l(k)$. Introduce the vector $c$ of unknowns with its components
\begin{equation}
    c_n=\left\{\begin{array}{ll}
        b_{nl}(k) & \text{for }n=0,1,\dots,v-1, \\
        \tan\delta_l(k) & \text{for }n=v.
    \end{array}\right.
\end{equation}
To find the unknowns $c_n$, consider $(v+1)$ equations
\begin{equation}
    \langle\bar\beta_{nl} | \widetilde{H}^l-E | \widetilde{u}_l\rangle = 0, \quad n=0,\dots,v.
    \label{Efros_eq}
\end{equation}
Here $\bar\beta_{nl}(r)$ 
comprise a
set of $(v+1)$ linearly independent SRFs. Following 
Ref.~\refcite{Sharaf2025},
we 
choose the first $v$ functions as $\bar\beta_{nl}(r) = \beta_{nl}(r)$ ($n=0,\dots,v-1$). The last function $\bar\beta_{vl}(r)$ is arbitrarily chosen so that it satisfies the requirement of linear independence
with the rest $\bar\beta_{nl}(r)$ with $n\leq v-1$. 
Eq.~\eqref{Efros_eq} can be written in matrix form,
\begin{equation}
    Ac=B,
    \label{Efros_matr_eq}
\end{equation}
where matrix $A$ and vector $B$ have the following matrix elements and components,
\begin{subequations}
    \label{A_Efros}
    \begin{alignat}{2}
        A_{nn'}&=\langle \bar{\beta}_{nl} | \widetilde{H}^l-E | \beta_{n'l} \rangle, &\quad &\left\{\begin{array}{ll}
            n=0,\dots,v,  \\
            n'=0,\dots,v-1,  
        \end{array} \right. \\
        A_{nv}&=\langle \bar{\beta}_{nl} | \widetilde{H}^l-E | \widetilde{G}_l \rangle, &\quad  &n=0,\dots,v;
    \end{alignat}
\end{subequations}
\begin{equation}
    B_n=-\langle \bar{\beta}_{nl} | \widetilde{H}^l-E | F_l \rangle, \quad n=0,\dots,v.
    \label{B_Efros}
\end{equation}
Solving the matrix equation~\eqref{Efros_matr_eq}, we obtain $\tan\delta_l(k)$ and unknown coefficients~$b_{nl}(k)$.


As for the choice of 
SRFs
$\beta_{nl}(r)$, it is natural to use either the set of $v$ oscillator functions $\varphi_{n'l}(r)$ ($n' \leq N$), or the set of $v$ eigenfunctions $\gamma_\lambda(r)$ of 
truncated Hamiltonian $\widetilde{H}^{tr,\,l}$.\cite{Sharaf2025} Here, the Hamiltonian $\widetilde{H}^{tr,\,l}$ is defined by its $(N+1)\times(N+1)$ matrix $\widetilde{H}_{nn'}^{tr,\,l} = \widetilde{H}_{nn'}^l$ ($n,n'=0\dots,N$) and eigenfunctions $\gamma_\lambda(r)$ satisfy the following equations,
\begin{equation}
    \widetilde{H}^{tr,\,l} |\gamma_\lambda\rangle = E_\lambda |\gamma_\lambda \rangle, \quad \lambda=0,\dots,N,
\end{equation}
\begin{equation}
    \langle\gamma_\lambda|\gamma_{\lambda'}\rangle = \delta_{\lambda\lambda'}, \quad \lambda,\lambda'=0,\dots,N,
\end{equation}
where $E_\lambda$ are the eigenenergies corresponding to eigenfunctions $\gamma_\lambda(r)$. The eigenfunctions are convenient to be ordered so that lower indices $\lambda$ correspond to lower eigenenergies $E_\lambda$, i.\,e., $E_\lambda \leq E_{\lambda'}$ if $\lambda < \lambda'$. 
The set of eigenfunctions is especially convenient for many-body applications in combination of the Efros method with NCSM
since otherwise one needs to construct the SRFs~$\beta_{nl}(r)$ as non-excited center-of-mass superpositions 
with the definite value of the total angular momentum~$J$ of
the NCSM basis functions which are  Slater determinants that are not characterized by a definite value of the total
angular momentum~$J$ and may have contributions of the center-of-mass excitations.
However, generally, 
one can 
define
the set of SRFs~$\beta_{nl}(r)$ as 
any
set of $v$ linearly independent superpositions of 
short-range functions.
As for the function $\bar\beta_{vl}(r)$, it was 
shown
in Ref.~\refcite{Sharaf2025} that 
it is reasonable to define it as
the first oscillator function outside the truncation boundary $N$, i.\,e., $\bar\beta_{vl}(r) = \varphi_{N+1,\,l}(r)$. Such a choice is made in order to provide a necessary coupling of the interaction ($n$ and $n' \leq N$) and the continuum ($n$ or $n'>N$) regions of the Hamiltonian matrix $\widetilde{H}_{nn'}^l$.

It is noteworthy that 
the modified
Efros method 
of Ref.~\refcite{Sharaf2025} 
is equivalent to 
the
HORSE approach if ${v=N+1}$. However, the advantage of
the 
Efros method is that one can obtain adequate results which converge to numerical solution 
if ${v<N+1}$
and in many cases even if~$v\ll N+1$. 
Besides, matrix elements $A_{nn'}$ and vector components $B_n$ are readily obtained if one uses expressions~\eqref{Sn_Cn} for 
the
Coulomb functions $F_l(\eta,kr)$ and $\widetilde{G}_l(\eta,kr)$ with coefficients $S_{nl}(k)$ and $C_{nl}(k)$ 
that are easily
numerically calculated as discussed in Sec.~\ref{Coul_scat}. Using $\varphi_{n'l}(r)$ or $\gamma_\lambda(r)$ as SRFs $\bar\beta_{nl}(r)$, one also obtains finite sums in the 
right-hand
sides of expressions~\eqref{A_Efros} and~\eqref{B_Efros}. Thus, Efros method provides a simple and numerically efficient approach to finding phase shifts $\delta_l(k)$
and  wave functions in scattering  of charged particles.

\section{Results}
\label{results}

We use 
the Woods--Saxon potential
as the nuclear potential,
\begin{equation}
V^{Nucl,\,l} 
                = \frac{V_0}{1 + \exp{\left( \frac{r-R_0}{\alpha_0} \right)}} + (\mathbf{l} \cdot \mathbf{s}) \frac{1}{r} \frac{d}{dr} \frac{V_{ls}}{1 + \exp{\left( \frac{r-R_1}{\alpha_1} \right)}},
    \label{WS}
\end{equation}
where $\mathbf{l}$ and $\mathbf{s}$ are the orbital momentum and the spin, respectively. To improve 
the
convergence,
as it was shown in Ref.~\refcite{YANIKOV2025100075}, it is important to use smothing\cite{JRevai_1985} of the
truncated 
matrix of the
nuclear interaction~\eqref{trunc_pot}, that is~$ \widetilde{V}_{nn'}^{Nucl,\, l}$ is replaced by
\begin{equation}
    \mathbb{V}_{nn'}^{Nucl,\, l}=\left\{\begin{array}{ll}
        \sigma_n V_{nn'}^{Nucl,\, l} \sigma_{n'} & \text{for $n$ and $n' \le N$}, \\
        0 & \text{for $n$ or $n'>N$},
    \end{array}\right .
    \label{trunc_pot_sm}
\end{equation}
where
\begin{equation}
    \sigma_n = \frac{1-\exp{\left(-{\left[\frac{\alpha(n-N-1)}{N+1}\right]}^2\right)}}{1-\exp{(-\alpha^2)}}.
\end{equation}
In what follows, the smoothing parameter $\alpha=5$ is used. 
Note, the Coulomb interaction~$\widetilde{V}_{nn'}^{Coul,\, l}$ [see Eq.~\eqref{trunc_pot_coul}] should not be smoothed.
Following 
the
Efros formalism, we calculate 
the
phase shifts $\delta_l(k)$ and corresponding partial
elastic scattering
 cross sections
\begin{equation}
    \sigma_l(k) = \frac{4\pi}{k^2}(2l+1)\sin^2\delta_l(k)  ,
    \label{cross_section}
\end{equation}
and 
compare the results with 
those
obtained within HORSE method for Coulomb scattering discussed in Ref.~\refcite{YANIKOV2025100075} as well as with the results found by numerical integration of Schr{\"o}dinger equation using Numerov method which 
will
be referred 
to as exact.

\begin{figure}[t]
    \centering
    \includegraphics[width=.98\textwidth]{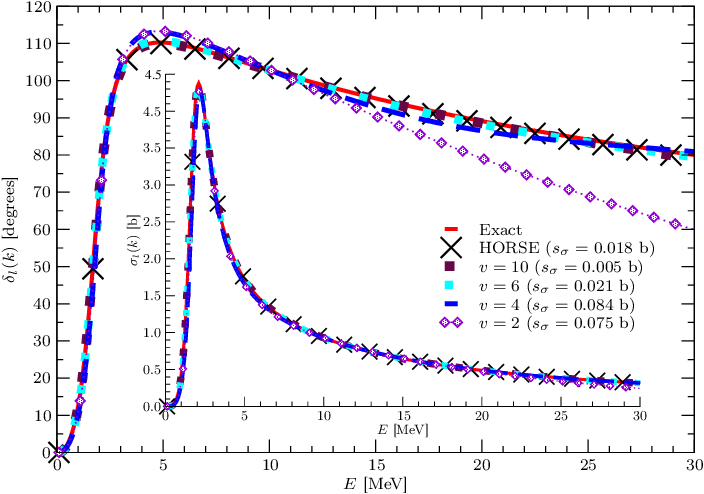}
    \caption{Phase shift $\delta_l(k)$ and 
 elastic $p{-}\alpha$ scattering   
        cross section $\sigma_l(k)$ 
 dependences  
         on relative motion energy $E$ 
        in the $p_{3/2}$
 partial       
        wave obtained by different methods. Solid line: numerical integration of Schr{\"o}dinger equation by Numerov method; crosses: HORSE method\cite{YANIKOV2025100075}; other lines: Efros method with different $v$ for SRFs $\bar\beta_{nl}(r) = \bar\beta_{nl}^{Eig}(r)$. For each method, $s_\sigma$ is the cross-section RMS deviation from the exact values. The results were obtained for truncation boundary $N=10$, 
    $\hbar\omega = 27$~MeV and starting number  $n_s = 200$
  for calculating $S_{nl}(k)$ and $C_{nl}(k)$   using Eq.~\eqref{TRR}.}
    \label{figure_1}
\end{figure}

Let us consider
the 
$p {-} \alpha$ 
scattering phase shifts
in the  $p_{3/2}$ partial wave.
We use the 
Woods--Saxon
potential parameterization suggested in Ref.~\refcite{BANG1979119} with ${V_0=-43.0}$~MeV, $R_0=2.0$~fm, $\alpha_0=0.70$~fm, $V_{ls}=40.0 \text{ MeV} \cdot \text{fm}^2$, $R_1=1.50$~fm, and $\alpha_1=0.35$~fm. The calculations of phase shifts were performed for different 
numbers of SRFs
$v$ and various sets of SRFs 
$\beta_{nl}(r)$ and 
 $\bar\beta_{nl}(r)$ 
(note, we always choose the first $v$ functions  $\bar\beta_{nl}(r)$ the same as $\beta_{nl}(r)$:
$\bar\beta_{nl}(r) = \beta_{nl}(r)$ for $n=0,\dots,v-1$)
 at oscillator energy $\hbar\omega = 27$~MeV which yields the best results. We started with choosing the set of SRFs $\bar\beta_{nl}(r)$ as
\begin{equation}
    \bar\beta_{nl}^{Eig}(r) = \left\{\begin{array}{lll}
        \gamma_{n}(r), & \quad & n=0,\dots,v-1, \\
        \varphi_{N+1,\,l}(r), & \quad & n=v.
    \end{array}\right .
    \label{beta_eig}
\end{equation}
We note that the same set of SRFs $\bar\beta_{nl}(r)$ was used in Ref.~\refcite{Sharaf2025} for 
the scattering of non-charged particles.
The results are shown in Fig.~\ref{figure_1}. One can 
see
that 
the
Efros method provides adequate phase shifts which converge to their exact values even for small $v=4$. It seems, however, that choosing $v=2$ leads to unsatisfying results for ${E>15}$~MeV. 
This is not surprising since the second (first excited) eigenstate in this case has an energy of 10.25~MeV; hence one
cannot expect an adequate description of the phase shifts at higher energies using only 2 SRFs and the phase shifts start 
deviating from the exact ones starting from approximately 11~MeV.
Nevertheless, the 
elastic
cross section $\sigma_l(k)$,
which, contrary to the phase shifts, is
an observable,
does not differ much from exact values since $\sin^2\delta_l(k) \sim 1$ in this region of energies 
[see Eq.~\eqref{cross_section}].
The difference from the exact solution can be actually expressed numerically 
by calculating the
root-mean-square (RMS) deviation $s_\sigma$ 
of the calculated cross section from the exact one
which appears to be quite small for all choices for $v$ (see the inset 
in
Fig.~\ref{figure_1}).

\begin{figure}[t]
    \centering
    \includegraphics[width=.95\textwidth]{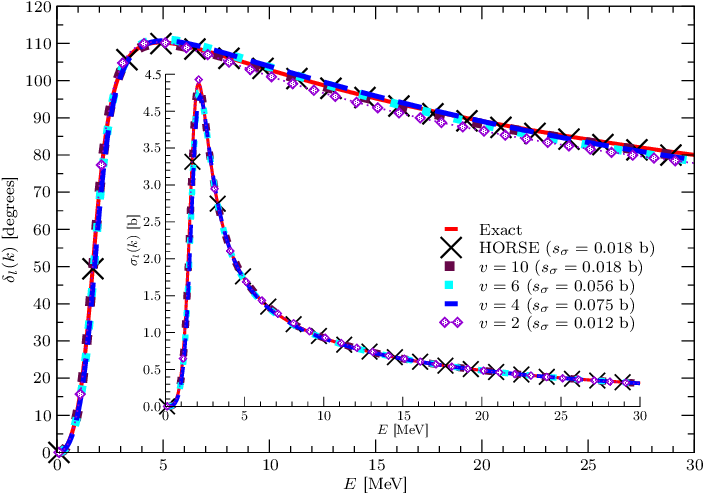}
    \vspace{-2.ex}  
    \caption{Same as Fig.~\ref{figure_1} but for SRFs $\bar\beta_{nl}(r)=\bar\beta_{nl}^{Osc}(r)$.}
    \label{figure_2}
%
    \centering
    \includegraphics[width=.95\textwidth]{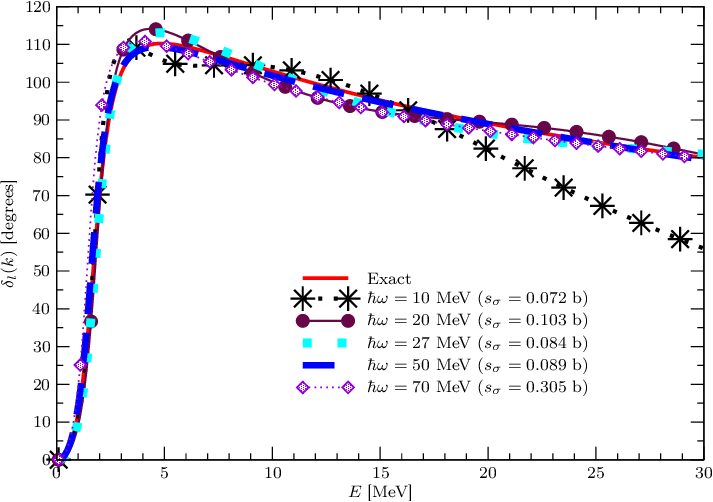}
\vspace{-1.5ex}    
    \caption{Same as Fig.~\ref{figure_1} but the results of the Efros method are obtained for $v=6$ with different $\hbar\omega$ values.
    }
    \label{figure_3}
\end{figure}

One
can also use the first $(v+1)$ oscillator functions $\varphi_{nl}(r)$ as the set of SRFs $\bar\beta_{nl}(r)$, i.\,e.,
\begin{equation}
    \bar\beta_{nl}^{Osc}(r) = \varphi_{nl}(r), \quad n=0,\dots,v.
    \label{beta_osc}
\end{equation}
The results obtained in this case (see Fig.~\ref{figure_2}) demonstrate also a good convergence. It is interesting that the use
of the oscillator functions as SRFs provide closer results to the exact ones (smaller RMS deviations) in the case of small~$v$
but the eigenfunction SRFs describe the exact cross sections better in the case of larger~$v$ values.

One can observe convergence of Efros method results with $\bar\beta_{nl}(r) = \bar\beta_{nl}^{Eig}(r)$ to exact solution for $\hbar\omega$ in the wide range from ${\sim }25$ MeV to ${\sim }50$ MeV if $v \gtrsim 4$ (see Fig.~\ref{figure_3}). For small $v \sim 2$, however, the method becomes more sensitive to
the $\hbar\omega$ variation.

\begin{figure}[t]
    \centering
    \includegraphics[width=\textwidth]{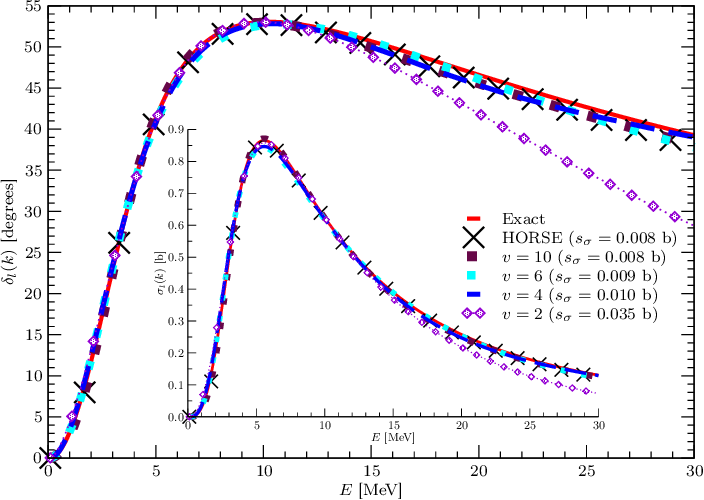}
    \caption{Same as Fig.~\ref{figure_1} but for 
$p_{1/2}$ 
  wave and $\hbar\omega = 22$ MeV.}
    \label{figure_4}
\end{figure}

It is useful to study other types of scattering with broader or narrower resonances. For example, one can consider 
the
$p_{1/2}$ 
partial
wave in $p{-}\alpha$ scattering with a 
wider
resonance (see Fig.~\ref{figure_4}). In this case, 
the 
Efros method provides a 
satisfactory
convergence to
the 
exact values if $v \ge 4$. 
Note, with $N=10$ and $\hbar\omega=22$~MeV, the second state in this case has an eigenenergy of 10.99~MeV, and in the case
$v=2$, the 
phase shifts and cross section 
start deviating at this energy from the exact results.

\begin{figure}[t]
    \centering
    \includegraphics[width=\textwidth]{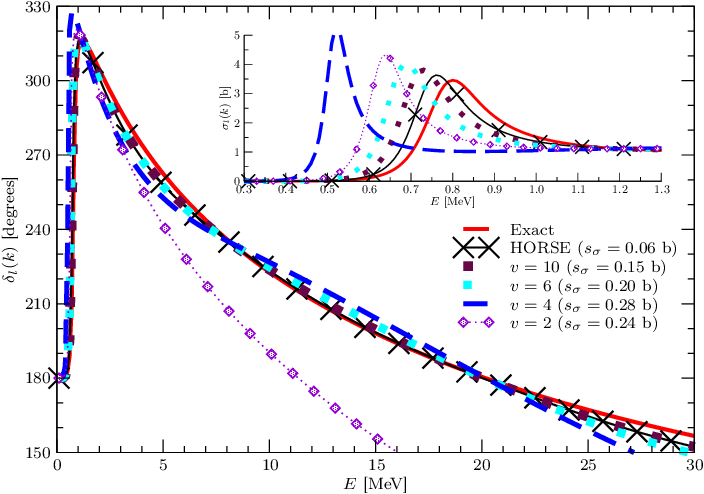}
    \caption{Same as Fig.~\ref{figure_1} but obtained in the $s_{1/2}$ wave for $p{-}^{15}N$ scattering with $\hbar\omega=18$ MeV.}
    \label{figure_5}
\end{figure}

To analyze accuracy of Efros method in case of 
different charges and a narrower
resonance, we 
examined the
$s_{1/2}$ wave in $p{-}^{15}N$ scattering using the parameterization of 
the Woods--Saxon
potential as in Ref.~\refcite{Bang1999PmatrixAJ}, i.\,e., $V_0=-55.91$~MeV, $R_0=3.083$~fm, $\alpha_0=0.53$~fm, $V_{ls}=0.9 \text{ MeV} \cdot \text{fm}^2$, $R_1=3.083$~fm, and $\alpha_1=0.53$~fm (see Fig.~\ref{figure_5}).
Though the phase shift 
seems to converge well to exact values for $v \ge 4$, in the region of small energies, the resonance position and its width value are 
sensible to the choice of $v$. It
is interesting
that using a small 
number of SRFs 
$v=2$, the position of the resonance is closer to its numerical value than 
in the case of
$v=4$, however, the phase 
shifts obtained with
$v=2$ 
start deviating from the exact ones at
$E \ge 5$ MeV. To improve the description of this resonance, one needs to increase the truncation boundary of the
potential energy matrix~$N$.

\section{Conclusion}
\label{conclusion}

In this 
paper, we 
developed the description of scattering of charged particles within the version of
the Efros method utilizing the expansion of the scattering wave functions in oscillator functions.
Using model problems which can be exactly solved numerically by other methods, we demonstrated that the proposed approach
provides accurate results for scattering phase shifts and cross sections even when only a relatively small number of short-range 
functions is used. 
As a result, we obtained
a promising tool for the 
{\it ab initio}
description of reactions with charged particles 
based on the NCSM calculations.

\section*{Acknowledgements}
This work is supported by the Russian Science Foundation under the grant No 24-22-00276.

\section*{ORCID}

\noindent Ustin M. Yanikov - \url{https://orcid.org/0009-0006-0532-7696}

\noindent Vasily A. Kulikov - \url{https://orcid.org/0000-0002-2770-3465}

\noindent Andrey M. Shirokov - \url{https://orcid.org/0000-0002-0331-6209}

\appendix

\bibliographystyle{ws-ijmpe}
\bibliography{references}

\end{document}